%% file: main.tex
\documentclass[sigconf,nonacm=true,9pt]{acmart}

\usepackage{subfiles}
\usepackage{graphicx}
\usepackage{adjustbox}
\usepackage{listings}
\usepackage{float}
\usepackage{color}
\usepackage{subcaption}
\usepackage{outlines}
\usepackage{listings}
\usepackage{xcolor}
\usepackage{enumitem}

\definecolor{codegreen}{rgb}{0,0.6,0}
\definecolor{codegray}{rgb}{0.5,0.5,0.5}
\definecolor{codepurple}{rgb}{0.58,0,0.82}
\definecolor{backcolour}{rgb}{0.95,0.95,0.92}

\lstdefinestyle{mystyle}{
    backgroundcolor=\color{backcolour},   
    commentstyle=\color{codegreen},
    keywordstyle=\color{magenta},
    numberstyle=\tiny\color{codegray},
    stringstyle=\color{codepurple},
    basicstyle=\ttfamily\footnotesize,
    breakatwhitespace=false,         
    breaklines=true,                 
    captionpos=b,                    
    keepspaces=true,                 
    numbers=left,                    
    numbersep=5pt,                  
    showspaces=false,                
    showstringspaces=false,
    showtabs=false,                  
    tabsize=1
}

\graphicspath{ {./images/} }
\usepackage{caption}
\AtBeginDocument{%
  \providecommand\BibTeX{{%
    \normalfont B\kern-0.5em{\scshape i\kern-0.25em b}\kern-0.8em\TeX}}}
\setcopyright{none}
\begin{document}
\sloppy
\title{Information Batteries}
\subtitle{Storing Opportunity Power with Speculative Execution}

\author{Jennifer Switzer}
\affiliation{%
  \institution{UC San Diego}
  \city{San Diego}
  \state{California}
  \country{USA}
}
\email{jfswitze@ucsd.edu}

\author{Barath Raghavan}
\affiliation{%
  \institution{USC}
  \city{Los Angeles}
  \state{California}
  \country{USA}
}
\email{barathra@usc.edu}

\input{abstract}

\maketitle
\pagestyle{plain}

\input{introduction.tex}

\input{background.tex}
\input{analysis.tex}
\input{design.tex}
\input{implementation.tex}
\input{evaluation.tex}
\input{discussion.tex}

\bibliographystyle{ACM-Reference-Format}
\bibliography{main}

\end{document}

%% file: abstract.tex
\begin{abstract}
Coping with the intermittency of renewables is a fundamental challenge, with load shifting and grid-scale storage as key responses. 
We propose Information Batteries (IB), in which energy is stored in the form of information---specifically, the results of completed computational tasks. Information Batteries thus provide storage through speculative load shifting, anticipating computation that will be performed in the future.

We take a distributed systems perspective, and evaluate the extent to which an IB storage system can be made practical through augmentation of compiler toolchains, key-value stores, and other important elements in modern hyper-scale compute.  In particular, we implement one specific IB prototype by augmenting the Rust compiler to enable transparent function-level precomputation and caching. We evaluate the overheads this imposes, along with macro-level job prediction and power prediction.
We also evaluate the space of operation for an IB system, to identify the best case efficiency of any IB system for a given power and compute regime.
\end{abstract}

%% file: introduction.tex
\section{Introduction}
\begin{figure}
    \includegraphics[width=\columnwidth]{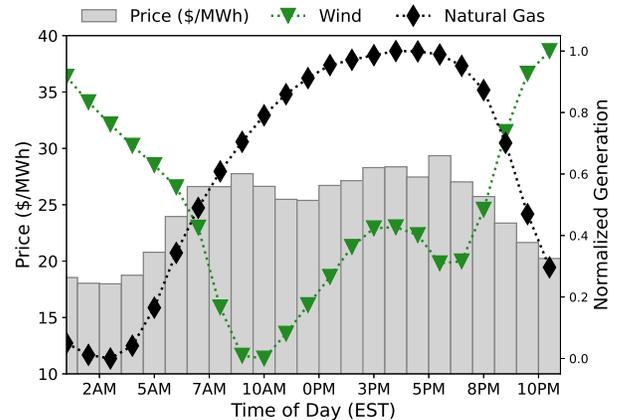}
    \caption{Gas generation is flexible; wind generation requires wind, and is often out of phase with demand. Prices tend to be lower when wind generation is higher. Based on 2019 price and fuel mix data from MISO~\cite{miso_data}.}
    \label{fig:motivation}
    \vskip -1.5em
\end{figure}

Within the next twenty years humanity must eliminate fossil fuels to avoid dangerous climate change~\cite{climatechange, ipcc}.  To do so requires renewable electricity generation.  While there have been dramatic decreases in the cost of wind and solar, their intermittency in response to weather and solar irradiance is a widely known issue~\cite{miso, caiso}. This intermittency is often out of phase with overall demands, so when renewable production is high, prices tend to be low (Figure~\ref{fig:motivation}). 

The research community has explored myriad responses to this intermittency~\cite{flex1,barker2012smartcap,johnson2011energy,mishra2012smartcharge,srikantha2012analysis,taneja2010towards,hegde2011optimal,xu2013incentivizing,mishra2013greencharge,correa2020s}, but has been stuck between the Scylla of dynamic load shifting and the Charybdis of expensive grid-scale storage.   Dynamic load shifting at grid scale requires the rare combination of \emph{flexible}, \emph{large}, and \emph{ubiquitous} loads.  Grid-scale storage with proven technology requires nearby hydroelectric capacity or expensive battery arrays.

In this paper we aim for the best of both worlds: storage of surplus renewable production through the load shifting of computation with \textbf{\emph{speculative execution}}.  Computation, unlike many other loads, is near-infinitely divisible, flexible, large, ubiquitous, and can be stored cheaply. This approach, \textbf{Information Batteries} (IB), entails storing energy as completed precomputations that can, as we show, meet or exceed the end-to-end efficiency of grid-scale storage \emph{using existing infrastructure}.  Not all workloads or conditions will yield high efficiency with this approach, so a key aspect of our exploration is its limits.

\vspace{0.5ex}
\noindent \textbf{Background.}
Even with relatively modest adoption of such renewable sources of power and despite the inherent statistical multiplexing of large power grids, grid operators will soon face the problem of too much power. For example, during the middle of the day in California it is now often the case that there is too much power being produced, largely due to solar, driving the price of electricity \emph{negative}~\cite{caiso}. This problem has arisen with just 20\% of California's electricity generation coming from solar (the renewable source with the greatest growth potential) \cite{cal_solar}. 

Today we face both load shedding due to insufficient power during critical times \emph{and} power dumping.  This disconnect in supply and demand, as renewable energy sources under-produce during low availability times and over-produce during high availability~\cite{greenware, renewableeconomics}, is the key challenge that must be solved to adopt renewables.\footnote{This negative-priced power and curtailed power (dumped power) are together referred to as \textit{opportunity power}~\cite{miso}.}

An oft-proposed strategy to address this problem is to increase grid-scale storage via traditional energy storage systems such as lithium-ion batteries and pumped hydro~\cite{storage}. However, such systems require a relatively high initial investment and have siting constraints; also, adding enough storage to soak up all excess production would quickly become prohibitively expensive~\cite{miso}. Similarly, smart-grid advocates often observe that if the grid could simply signal to individual devices, such as appliances, when to consume power, this demand shifting could adapt to power availability.  However, as with storage, demand shifting requires both grid upgrades---a proposition that a number of governments have balked at---and wide-scale adoption of new smart demands.

\vspace{0.5ex}
\noindent \textbf{Approach.}
With Information Batteries, we propose the storage of energy as information, using the large and growing footprint of computing to perform both the functions of storage and load shifting.  This approach hinges on three observations. First, data centers worldwide consume large amounts of electricity (250--500 TWh in 2018) and are projected to become even more power-hungry (840--3640 TWh in 2030~\cite{datacenteruse}).\footnote{While some companies, such as Google, balance their data center power usage with renewable energy power purchase agreements, this still addresses only \emph{average} power generation, not the peaks and troughs of generation.}  Second, many computational tasks can be precomputed in whole or in part.  Third, both power availability \emph{and} compute demands are somewhat predictable, and thus it is possible to do \emph{speculative} load shifting.

Rather than storing excess energy as a chemical (lithium-ion) or gravitational (pumped hydro) potential, an IB system stores this as information---completed computations. When excess renewable energy is available, an Information Battery system uses this excess to perform precomputable, energy-intensive computations. The results of these computations are then stored for when they are needed.  The duality of computation and energy is not new---it has been studied extensively in information theory, such as in the context of adiabatic computing~\cite{denker1994review}.  However, this observation has not been raised to the level of grid-scale energy systems.

We study \emph{the limits of Information Batteries} in enabling a renewable energy transition at a macro scale. For Information Batteries to be effective and worthwhile, they must 1) be less expensive than traditional storage systems, 2) shift significant computational work from grid to opportunity power, and 3) achieve this significant shift for some common workloads rather than new, esoteric workloads. As we show, it seems that this is only possible for some workloads and in some contexts.  We evaluate the space of operation for Information Batteries to identify the best case efficiency of \emph{any} IB system for a given power and compute regime.

We take a distributed-systems perspective, and evaluate the extent to which an IB storage system can be made practical through augmentation of compiler toolchains, key-value stores, and other important elements in modern hyper-scale compute.
Our ability to shift computational load to opportunity power hinges on the accuracy of our predictive engine. If we cannot predict upcoming requests with at least reasonable accuracy, our system will not have results available for when they are requested, and may end up wasting opportunity power performing computations that are never requested. Furthermore, we must be able to accurately predict the future availability of opportunity power, so that we can effectively schedule our computations to take advantage of it. Finally, we must ensure that the cost of retrieving cached results is significantly smaller than the cost to compute them; otherwise, it is more efficient to perform the computation on demand.

\vspace{0.5ex}
\noindent \textbf{Contributions.} This paper makes three contributions:
\begin{enumerate}[leftmargin=*, topsep=2pt]
\item We introduce the idea of Information Batteries that provide a new speculative load shifting mechanism to address grid-scale renewable energy intermittency.

\item We explore the design space of Information Batteries and show that in some common power and compute regimes there is the potential for an IB system to deliver efficiencies better than the best grid-scale storage available while in others IB systems provide little benefit.

\item We implement a proof-of-concept IB system by augmenting the Rust compiler to enable transparent function-level precomputation and caching. We evaluate overhead, along with macro-level job prediction and power prediction.
\end{enumerate}

%% file: background.tex
\section{Context}
In this section we motivate the need for and feasibility of Information Batteries. We focus on two renewable energy markets: the Midcontinent ISO (MISO), which operates in the Southern and Midwestern United States and parts of Canada, and the California ISO (CAISO), which covers all of California. This allows us to narrow our scope while still considering both wind-dominant (MISO) and solar-dominant (CAISO) markets ~\cite{aboutmiso}.

\subsection{Availability of opportunity power}

Opportunity power in CAISO and MISO is significant, growing, and often available. Current estimates place the yearly combined opportunity power of CAISO and MISO in 2017 between 7--20 TWh per year~\cite{miso, caiso}. In MISO, opportunity power is available 99\% of the time (meaning opportunity power is available \textit{somewhere} 99\% of the time, since prices are location-dependent), and often in intervals of >100 hours~\cite{miso}. In CAISO, some solar generators experience 3.3 hours of opportunity power per day~\cite{caiso}.

Solar and wind energy are projected to be the fastest growing sources of electricity generation in the U.S., and currently account for ~10\% of total U.S. electricity generation~\cite{eia,c2es}.
As solar and wind generation grow, so too will the amount of curtailed and negative priced power. Indeed, \cite{caiso} measures the compound annual growth rate of opportunity power in CAISO to be 40\%. Assuming 1.5 TWh of opportunity power in CAISO in 2017 (a conservative estimate) and a constant growth rate, CAISO alone could provide 22 TWh of opportunity power by 2025, enough to power all of Los Angeles.

\subsection{Limitations of traditional energy storage}
Energy storage is a simple response to overproduction. However, current battery prices make this untenable; grid-scale lithium-ion storage costs \$209 per kWh today~\cite{batteryprices}, not including installation costs. A na\"ive analysis of CAISO and MISO data, assuming 1.5 TWh/year and 6 TWh/year of opportunity power, respectively, yields a conservative estimate of \$35 million to add one hour of storage to CAISO, and \$140 million to add one hour of storage to MISO.
A more complex analysis \cite{miso} suggests that adding grid-scale storage provides diminishing returns, and that adding 50 hours of storage to MISO would cost \$50--400M \textit{per wind generation site}, on par with the cost of the turbines themselves.

\subsection{Non-computational load shifting}
Existing non-computational flexible loads include manufacturing facilities, EV charging, and adaptive home appliances~\cite{flex1,flex2}.

Household or otherwise small-scale (but widespread) loads are a popular demand shifting target, such as in smart homes with or without storage~\cite{barker2012smartcap,johnson2011energy,mishra2012smartcharge,srikantha2012analysis,taneja2010towards,hegde2011optimal,xu2013incentivizing} or in smart buildings more generally~\cite{mishra2013greencharge}.  EV charging is a growing, flexible load~\cite{correa2020s}.
Such load shifting requires accurate prediction capabilities~\cite{demandbal,rnnweather} such as of renewable generation and weather~\cite{bashir2019solar,sharma2010cloudy,sharma2011predicting}.  

\subsection{Computational load shifting}
Some have considered to simply store power in data centers using the batteries in those facilities rather than shifting load~\cite{urgaonkar2011optimal,govindan2011benefits,goiri2013parasol}.  Prior work has suggested shifting the loads themselves to leverage surplus power~\cite{chien2019zero}, for example by examining the price for computation (or power itself) in different regions~\cite{qureshi2009cutting,liu2013data}.

Data centers have more than enough capacity to soak up opportunity power. American data centers consume 70TWh/year, accounting for 1.8\% of the country’s total energy consumption~\cite{enreport}. With opportunity power in CAISO and MISO estimated at 7--20TWh/year \cite{miso, caiso}, opportunity power has the potential to provide between 10--30\% of the energy needed by data centers. 

Prior work also considered scheduling large-scale compute tasks under a variety of constraints. 
Speculative execution has of course also been long known in computing~\cite{chang1999automatic,smith1990boosting}, but has not been applied in this context.
Google recently announced their ``carbon-intelligent computing platform'' that attempts to reduce carbon emission of datacenters by aligning time-insensitive tasks with periods of high renewable energy availability~\cite{google}. In this paper, we will refer to this and similar techniques as time-shifted compute.

%% file: analysis.tex
\begin{figure*}
    \centering
    \includegraphics[width=0.8\textwidth]{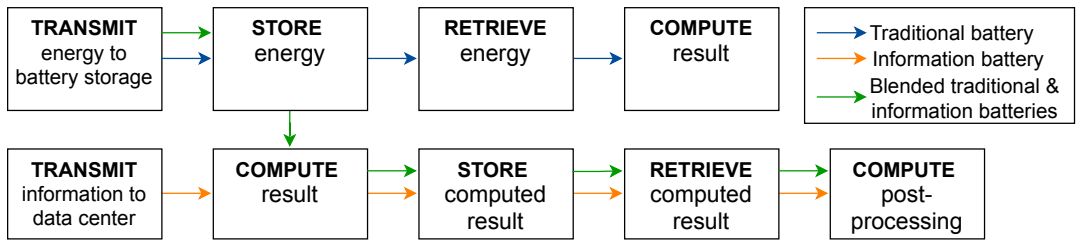}
    \vskip 0.1em
    \caption{Energy flows differently through Information Batteries and traditional batteries.}
    \label{fig:eflow}
    \vskip -0.75em
\end{figure*}

\section{Information Batteries}
Next we address the challenges of: 1) properties of computational tasks, hardware, and energy grids that make them more or less suited to Information Batteries, 2) the key elements of an IB system, and 3) the best case scenario/theoretical limits of IB systems.

\subsection{Prediction}
There is an element of speculation to any demand shifting: the assumption that the work done will be useful at some future time. However, most flexible loads are general tasks that will almost certainly be useful (e.g., charging an EV, washing clothes), and are therefore non-speculative.
Precomputation, on the other hand, is speculative. Given the infinite space of possible computations, more work is needed to identify computational work that will be useful at some point in the future. The ability to predict future computations---to perform \emph{task prediction}---is therefore key.

\subsection{Granularity}
Prior work on flexible loads has focused on high-granularity tasks---washing clothes, charging a car, even time-shifting large-granularity computational tasks~\cite{google}. Micro demand shifting is comparatively understudied.
IB systems allow for tuneable load granularity, since computational tasks can (at least in theory) be broken into different-sized tasks (although data dependencies and complexity considerations may in practice favor large tasks). 

\subsection{Speculative demand shifting}
The IB approach of demand shifting is differs from prior approaches in two ways: the load itself is not pre-existing (as it is \emph{speculative}), and the granularity at which we shift varies.  Computation can be speculatively executed at many granularities, from whole-system to individual instructions.

Speculative load shifting uses energy to store not energy but \textit{information}.  This is different from conventional load shifting, which does not require storage (as the load itself has been shifted).  However, since data storage is far less expensive than energy storage, this requirement is a minor imposition.

\subsection{Computational loads}
Central to IB effectiveness is the type of computation available for execution.
To that end, we consider the effect of a task's run time and data dependence on IB effectiveness.

Information Batteries are well-suited to tasks with high predictability and a large potential speedup. Potential speedup is the ratio between run time and IB cache latency. Given the same cache latency, longer running tasks will have higher potential speedup. OpenAI notes that AI workloads, particularly training, are growing exponentially~\cite{openai-compute}. Such workloads are ideal for Information Batteries due to their size, latency insensitivity, and high predictability. Also, compute with less data dependency (of any size) is more predictable, as there are fewer possible inputs.

In addition to predictability of macro workloads, the IB approach can precompute some or all of the fragments of expected jobs and then reassemble these fragments on demand. Thus predictability extends not only to whole jobs but sub-job fragments. Furthermore, with careful binary and execution trace analysis, it is likely possible to perform computations using speculatively precomputed fragments of \emph{different} jobs, as many workloads have some commonality. 
Indeed, assembling whole computations out of fragments of code is well studied in the literature in very different contexts, such as in the case of Return-oriented Programming~\cite{shacham2007geometry,roemer2012return}.
Exploring the efficiency of fragment precomputation and reassembly is worthy of study but beyond the scope of this paper.

\subsection{Grid properties}
For an IB system to be successful, renewable generation must be large, long-lasting (e.g., with high duty cycles), and predictable. This is influenced by consumption patterns, the power mix of the grid, the physical locations of energy generators, and the amount of traditional (battery) storage available.

\subsubsection{Renewable energy availability}
The grid power mix impacts energy availability and pricing. Energy generation from fossil fuels can be scaled up or down to meet demand, hydro generation is relatively stable and flexible, and intermittent renewables (primarily wind and solar) produce according to environmental conditions. These different patterns have an impact on pricing: when generation is higher than demand, prices drop, as expected. 

\subsubsection{Power prices}
Negative power prices are the best indicator of uneconomic production. The frequency and duration of negative-priced power, its geographic spread, and the predictability of power prices all influence the ability of an IB system to identify and use uneconomic renewable supply.

\subsection{Framework}

The key idea behind Information Batteries is quite simple: when renewable energy is available, we use it to speculatively perform computation. The challenge is in determining \emph{what} computation to perform, \emph{where} and \emph{when}, and \emph{how} these computations should be done to make it efficient to retrieve their results later. The energy expended to perform the computation is therefore stored as the result of a computational task.\footnote{There is a long history of work in lower-level aspects of the relationship between energy and information, such as work on adiabatic computing~\cite{denker1994review}.}

%% file: design.tex
\section{Design}
Next we describe the key elements of an IB system, with a focus on several goals: to maximize the shift of compute to opportunity power (e.g., for a significant subset of common workloads) and to do so at a lower cost than traditional energy-storage systems.

\subsection{Overview}
Next we explore the components of an IB system, from compiler toolchains to ML-based tools for power and job prediction. We discuss how these could be combined in a production system.

\paragraph{Compiler Toolchain.} Application binaries must be augmented to allow for run-time caching and retrieval of results.
\paragraph{Key-Value Store.} The system must include an efficient key-value store for run-time caching and retrieval. In particular, the latency of precomputed result retrieval from the key-value store must be significantly less than the latency of computing that result directly. 

\paragraph{Job Scheduler.} A management system must be in place for scheduling incoming jobs during periods of opportunity power, and returning results once they are available. In particular, jobs must be prioritized according to their latency sensitivity and run time. The job scheduler must therefore be aware of the target workload.

\paragraph{Predictors.}
Information Batteries rely on speculative execution, and therefore require accurate predictions of future conditions. Specifically, the IB system must be able to predict: 1) whether or not opportunity power will be available in the near future (e.g., what the price of energy will be) and 2) what jobs will be requested in the near future, so that these jobs can be procomputed.

\begin{figure}
\centering
\footnotesize
\begin{lstlisting}[language=Python]
cycles_trad = 0
cycles_op = 0
every 5 minutes:
  predict = model price prediction
  actual = actual price according to trace
  job_size = number cycles per job
  hit_rate = accuracy of task prediction
  if predict<0 and actual<0: #true positive
    cycles_op += job_size
    cycles_trad += (1 - hit_rate) * job_size
  elif predict<0 and actual>0: #false positive
    cycles_trad += job_size
                + (1 - hit_rate) * job_size
  elif predict>0 and actual<0: #false negative
    cycles_trad += overhead of memoize
  else: #true negative
    cycles_trad += overhead of memoize
        
\end{lstlisting}
  \caption{Logic of the IB simulator.}
  \label{lst:sim}
  \vskip -2em
\end{figure}

\subsection{Program instrumentation}
Memoization is a well-known technique used to improve the performance of programs by caching the results of expensive computations for later use. Previous work has considered the benefits of program memoization at the trace level, the basic-block level, and the function level~\cite{bbmemoize, trace_level_reuse, dynamic_trace_memoization}. In this paper, we employ function-level memoization, but later propose generalizations to this.

Prior work has shown that a subset of functions can benefit from software memoization at compile and load time. These functions share the following characteristics: expensive, side-effect free, critical, and repetitive arguments~\cite{ltmemoize}. Compile-time memoization achieved modest performance gains on a subset of SPEC benchmarks that contain memoizable functions in the critical paths~\cite{ctmemoize}. 

We apply these memoization techniques to precompute results for memoizable functions in the program's critical path. These results obviate later execution with non-renewable power. We achieve this by instrumenting all application binaries to use a key-value store to memoize the results of function calls.

\subsubsection{Compiler extensions}
We consider one specific approach to program instrumentation---compiler extension---and highlight other possibilities later. In this approach, all source code is compiled using our customized compiler, which inserts the appropriate hooks to precompute and retrieve precomputed results.

For each call to a precomputable function, the compiler inserts a fetch instruction to first check whether the function has been precomputed, and return the precomputed result if it exists.
Note that since the fetch occurs at run time, it is important that its overall latency is significantly lower than the run time of the function to be computed. It is therefore important both that our cache implementation be efficient, and that our definition of precomputable function excludes those with extremely short runtimes.\footnote{The definition of \emph{precomputable function} is flexible, and may differ between workloads.}

\subsubsection{Caching infrastructure.}
The IB cache must enable fast caching and retrieval of precomputed results.  In addition, for deployability, we would like to use a generic key-value store for caching rather than a IB-specific system. Here, latency and hit rate are important because we do not wish cache retrieval to itself induce higher overheads than computation itself. In addition, the memory footprint of the cache is important; results that require too much storage will increase the cost of the cache infrastructure itself.

\subsection{Scheduling}
There are three main components to the IB manager: the scheduler, which receives computational tasks and schedules them to maximize grid power savings; the price predictive engine, which makes predictions about upcoming power prices, including when prices will drop below zero; and the precomputation engine, which makes predictions about upcoming tasks, and performs precomputations on these predicted tasks whenever opportunity (negative-priced) power is available. Negative-priced power is not required for Information Batteries to be useful, but provides a useful proxy for surplus renewable generation.

\textit{The scheduler} receives tasks, and determines when and how they should be computed. Submitted tasks are of the form \verb+source code, deadline+, where \verb+deadline+ indicates the latest timestamp at which the result is needed.

If opportunity power is currently available, the task can be compiled (if not already compiled) and computed immediately. Otherwise, the scheduler asks the solar predictive engine when opportunity power will next be available. If the next available window of opportunity power is within the task's deadline, the task is scheduled to take advantage of that window.

Note that even if a task is not precomputable, the scheduler will still attempt to schedule it for a period of opportunity power. Thus any task with a generous enough deadline can be scheduled to use opportunity power.
The scheduler also forwards the source code and the time it was received to the precomputation engine, which bases its predictions on this real-time stream of task requests.

\textit{The renewable predictive engine} forms the core of the scheduler; the scheduler's overall effectiveness is dependent on the ability of the solar predictive engine to correctly predict opportunity power. An inaccurate solar predictor risks missing out on opportunity power, or erroneously scheduling tasks during regular grid power. We base our implementation on existing techniques for weather prediction using Recurrent Neural Networks (RNNs)~\cite{rnn1}. 

\textit{The precomputation engine} is responsible for predicting upcoming tasks, pre-computing them, and caching the results. 
There are two main components of the precomputation engine: the task predictor, and the precompute manager.
The task predictor receives a continuous stream of task requests from the scheduler, and uses a recurrent neural network to predict task requests. 

The precompute manager consists of a single event loop that queries energy prices every 5 minutes until they fall below some very small threshold (which may be zero or even negative). At this point, the manager requests 5 minutes worth of computational tasks from the task predictor. The precomputation engine functions in five minute increments since that is the smallest granularity at which energy prices are set in CAISO and MISO. 

\subsection{Integration}
Information Batteries are designed to work with existing data centers. Some, very limited, processing power is reserved for the IB manager, which manages the scheduling of both real-time computational tasks and precomputation. A cluster of machines or VMs is designated for precomputation; these perform precomputations when opportunity power is available. The IB cache, which stores the results of these precomputations, is stored locally for quick retrieval, though more advanced caching systems could be envisioned for larger jobs. No additional infrastructure is needed.

Although Information Batteries have thus far been pegged as an alternative to traditional energy storage, it is also possible to use them in conjunction with traditional batteries. Figure~\ref{fig:eflow} illustrates the flow of energy in a traditional and information battery system, and how these two could be combined.

%% file: implementation.tex
\begin{figure}
 \begin{minipage}{\columnwidth}
  Before Instrumentation:
\begin{lstlisting}[language=C]
...
fn square(a: u32) -> u32 { a * a };
let result = square(x);
...
\end{lstlisting}
 \end{minipage}
 \begin{minipage}{\columnwidth}
 After Instrumentation:
\begin{lstlisting}[language=C]
...
fn square(a: u32) -> u32 { a * a };
let result = memoize(square, "square", x);
fn memoize(f: fn(usize) -> usize, fname: String, a: usize) -> usize {
    let cached = db.get(fname); 
    if cached.is_some() {
        return cached.unwrap();
    }
    else {
        let computed = f(x);
        db.put(fname,computed);
        return computed;
    }
}
\end{lstlisting}
 \end{minipage}
 \caption{Instrumented programs first check for precomputed, cached results and use those results if found.}
 \label{lst:rusty}
 \vskip -2em
\end{figure}

\section{Implementation} \label{implementation}
Next we describe our proof-of-concept implementation of Information Batteries, which has three key components: 1) a Rust compiler augmentation for function-level precomputation, 2) a price prediction model for both CAISO and MISO, and 3) a function-level task prediction model. 
Benchmarking these components allows us to realistically parameterize our IB simulator.

\subsection{Rust compiler instrumentation}
We augment the Rust compiler to do function-level precomputation. 
This is accomplished at the MIR (Mid-Level Intermediate Representation) stage of the Rust compiler~\cite{rust_book}. At this stage, the program has been converted into a control-flow graph (CFG) representation~\cite{rustc_book}. We implemented our instrumentation as an extra pass through the CFG, which we call the memoize pass. 

The memoize pass replaces each function call with a call to our \verb+memoize+ function\footnote{We require that the input binary include the definition of memoize}. This takes as input a pointer to the original function, the function's name (fetched at compile time) and the original arguments to the function. \verb+memoize+ does the following: 1) checks if the function has been called with the particular arguments before, and 2) executes and caches the result of executing the function if not. We refer to this instrumentation as the \verb+memoize+ wrapper. For simplicity, we support only functions with the function signature \verb+fn(u32)->u32+, but in practice other signatures could be supported.

Figure~\ref{lst:rusty} shows the instrumentation of a simple program with \verb+memoize+ (code simplified for readability).

\subsection{Cache}
The performance of the precomputation engine is highly dependent on the cache implementation. Latency should be minimized as much as possible. For our proof of concept, we use pickleDB-rs \cite{pickles} to implement a simple, local key-value store. However, any key-value store could be used in practice.

\subsection{Price Predictor}
We implemented our price models using TensorFlow, as Recurrent Neural Networks (RNN) with one LSTM layer and one dense layer. This is similar to techniques used for weather prediction.\cite{rnnweather,tf_weather}. 
We collected training data from historical 5-minute Location Marginal Prices (LMPs) reported by MISO~\cite{miso_data} and CAISO~\cite{caiso_data}. 

LMP is the dollar cost of supplying the next MW of power at a specific geographic region \cite{lmp}. Since pricing is location dependent, the model makes predictions based on time of day and geography. Given a location and a 5-minute interval, the price predictor returns a prediction for the next hour's worth of LMP prices. 

Although LMP prices do not directly reflect the amount of opportunity power in the grid, they provide a valuable proxy. Since most opportunity power consists of uneconomic, negative-priced power \cite{miso}, we define opportunity power as being available any time the LMP prediction drops below some small, tune-able threshold.

\subsection{Task Predictor}
We implemented task prediction using TensorFlow, as an RNN with one LSTM layer and one dense layer. We collected training and validation data from several popular open-source Rust crates: Substrate \cite{substrate}, Iced \cite{iced}, and Juniper \cite{juniper}. For each crate, we generated a function-level trace on their provided example applications, and used these traces to train the model.

The resulting model takes as input a series of function calls, and a specification, N, for how far in the future to predict. It returns a prediction for the next N function calls. This is similar to techniques used for text prediction \cite{tf_text}.

Note that this approach is quite simplistic. We do not consider, for instance, the value of arguments to the function. Our implementation is intended as a small proof-of-concept only.

%% file: evaluation.tex
\begin{table}
    \centering
    \begin{tabular}{lp{4cm}}
        \textbf{Parameter} & \textbf{Description} \\
        \hline
        Cache latency & Storage and retrieval overhead \\
        Cache hit rate & Accuracy of task predict model\\
        Price predict false positives &  \small{Fraction of time model mistakenly predicts negative-priced power}\\
        Price predict false negatives & Fraction of time model fails to predict negative-priced power\\
        \hline
    \end{tabular}
    \vskip 0.5em
    \caption{Input parameters for the system-level IB simulator.}
    \label{tab:sim}
    \vskip -1.5em
\end{table}

\section{Evaluation}
Our evaluation is in two parts. First, we microbenchmark each component of our proof-of-concept implementation described in Section~\ref{implementation}. We then use these microbenchmarks, combined with real price data from CAISO and MISO, to provide a realistic parameterization of a system-level simulation.

\subsection{Microbenchmarking}
Next we present our microbenchmarks of function-level memoization and price prediction.

\subsubsection{Function-level memoization.}
\begin{figure}
    \centering
    \includegraphics[width=0.9\columnwidth,height=50mm]{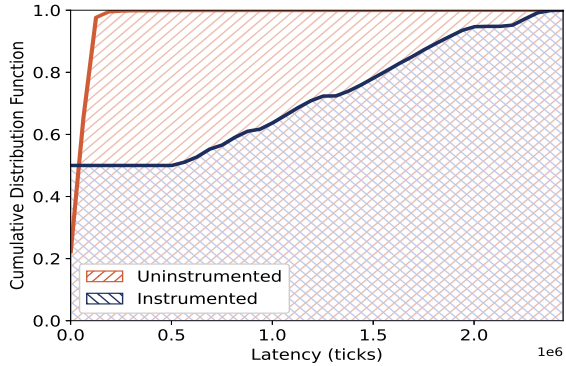}
    \caption{Function-level memoization with a 0.5 hit rate (meaning 50\% of tasks were precomputed, similar to the performance of our model). Cache misses take significantly longer in the information batteries scheme than they would if computed directly (due to the overhead incurred by added logic, and the reads and writes to the cache.) Cache hits return results significantly faster than traditional compute.}
    \label{fig:microbench}
    \vskip -1em
\end{figure}

Figure~\ref{fig:microbench} shows the latency of function calls with memoization, as compared to traditional compute. There are two sources of added latency from memoization: (1) The cost to check the cache for a precomputed result, and to return it if it exists; (2) The cost to store a computed result. 

We measured these values for our proof-of-concept implementation. The results, summarized in Table~\ref{tab:mem}, were used to parameterize our system-level simulations.
\begin{figure}
    \centering
    \includegraphics[width=0.9\columnwidth,height=50mm]{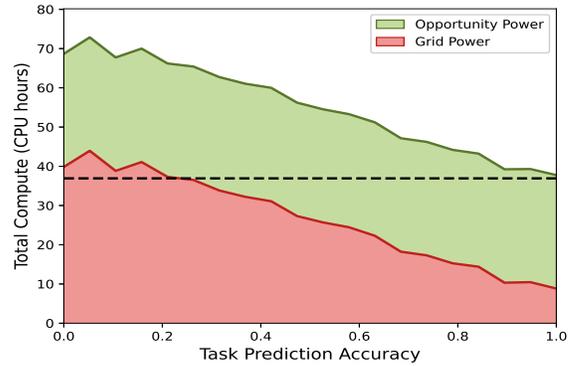}
    \caption{While more accurate task prediction increases performance, energy can still be saved with relatively low (e.g. 30\%) accuracies. Results are shown for machine cycles offloaded in simulation with varied task prediction accuracy and other parameters set based on microbenchmarking results. Simulation ran on MISO price data from 2019.}
    \label{fig:caching}
\end{figure}
\subsubsection{Price predictor}
\begin{figure}
\centering
\includegraphics[width=0.9\columnwidth]{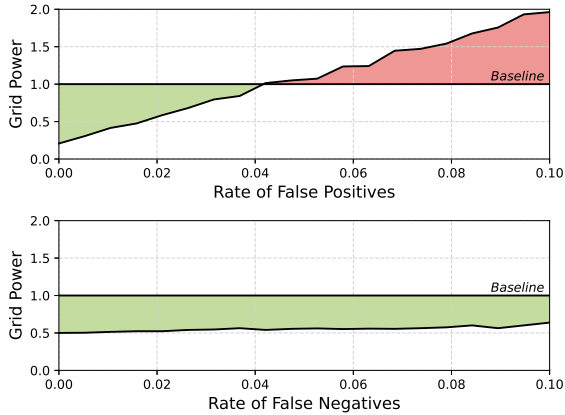}
\caption{Price predictor behavior. Low rates of false positives have a large impact on overall performance, since false positives cause the information batteries system to mistakenly schedule tasks for periods of low renewable production. False negatives have a significantly smaller effect.}
\label{fig:fp}
\vskip -1em
\end{figure}

Initial simulations indicated that the accuracy of the price predictor was extremely important to the overall performance of the information batteries system. However, false positives and false negatives had a very different effect; While keeping the rate of false positives low was extremely important (Figure~\ref{fig:fp}), false negatives had a relatively small impact. 

We measured the performance of the price predictor on both CAISO and MISO data. Table~\ref{tab:model_results} summarizes the results.

\subsubsection{Task predictor} Our naive task predictor had a top-1 accuracy prediction accuracy of 46\% for predicting the next 10 function calls given the previous 10. This means that each individual function in the predicted sequence has a 46\% chance of being correct.

\subsection{System-level simulation}
We built a system-level simulator to explore the performance of Information Batteries at scale. Our simulator takes as input a time-series of energy prices, and a set of parameters that define the performance of individual components of the system. We summarize these parameters in Table~\ref{tab:sim}.

The output of the simulator is a simulated 100-day run of an IB system, reporting 1) $\mathrm{cycles}_{\mathrm{avail}}$, the total amount of opportunity power that was theoretically available for compute, 2) $\mathrm{cycles}_{\mathrm{op}}$, the amount of opportunity power that was actually used for compute, and 3) $\mathrm{cycles}_{\mathrm{grid}}$, the amount of grid power that was used for compute. The run is considered successful if the amount of grid power used for compute is less than what would have otherwise been used in a traditional computing system.

For each five minute time interval, the simulator decides (based on the provided traces and the specified performance of the price predictor), whether or not to anticipate a period of negative-priced power. If negative-priced power is anticipated, it ``schedules'' jobs. It then records the amount of grid and opportunity power ``used'' in the interval. We summarize the logic of the simulator in Figure~\ref{lst:sim}.

\begin{figure*}
\centering
\subfloat[Small (100ms) Job]{\includegraphics[width=5cm]{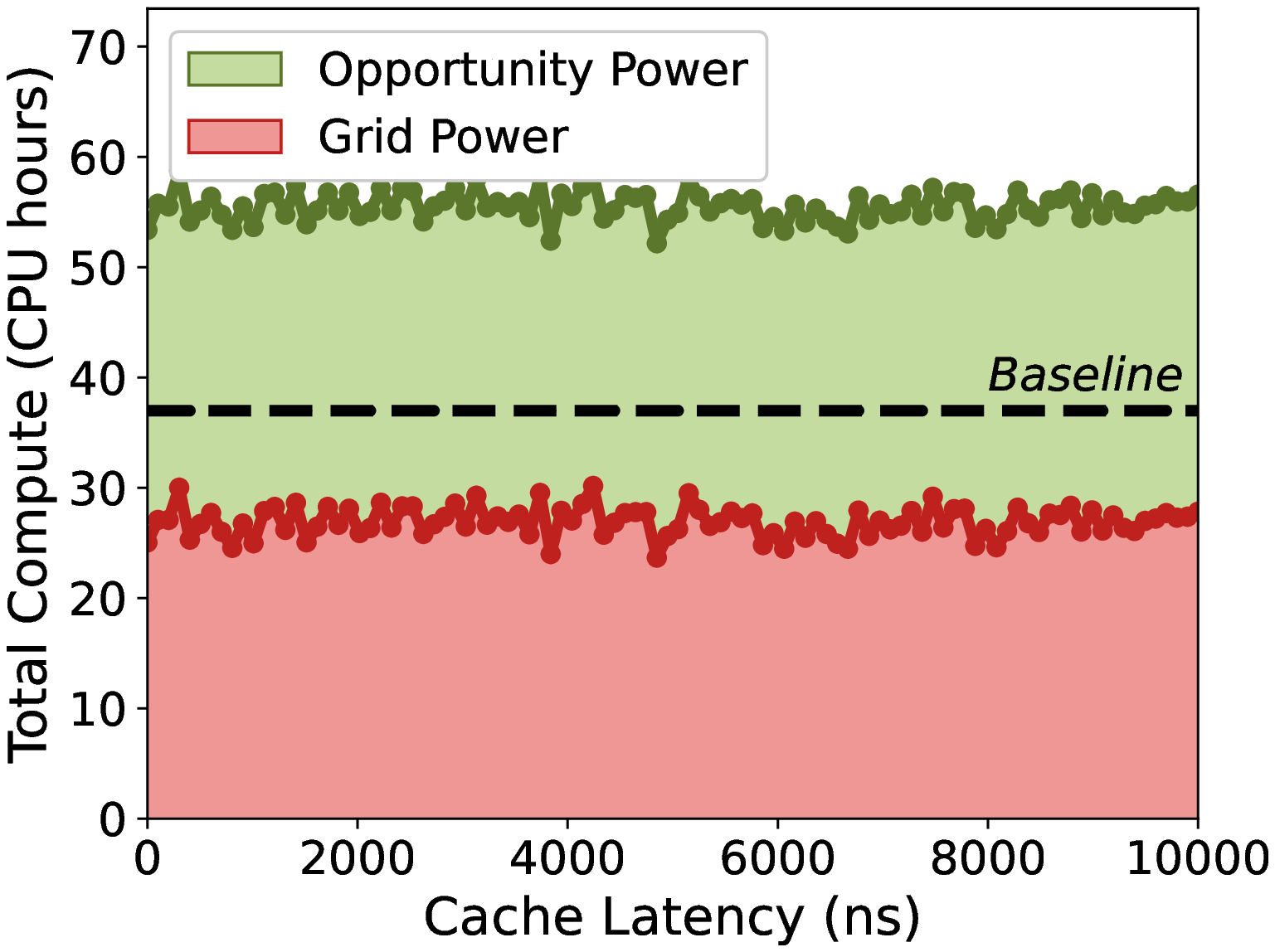}}\hfil
\subfloat[Medium (10s) Job]{\includegraphics[width=5cm]{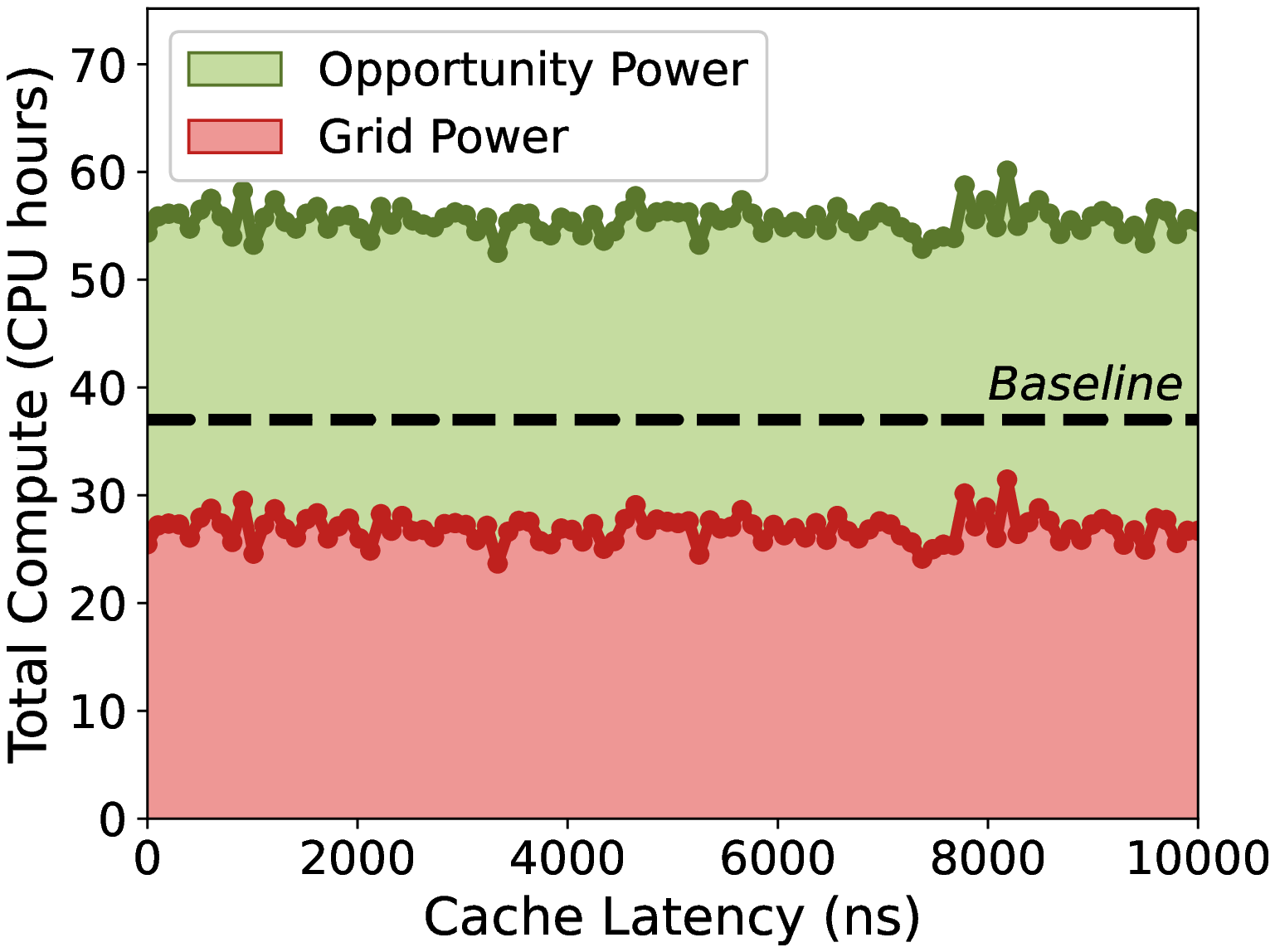}}\hfil 
\subfloat[Large (60s) Job]{\includegraphics[width=5cm]{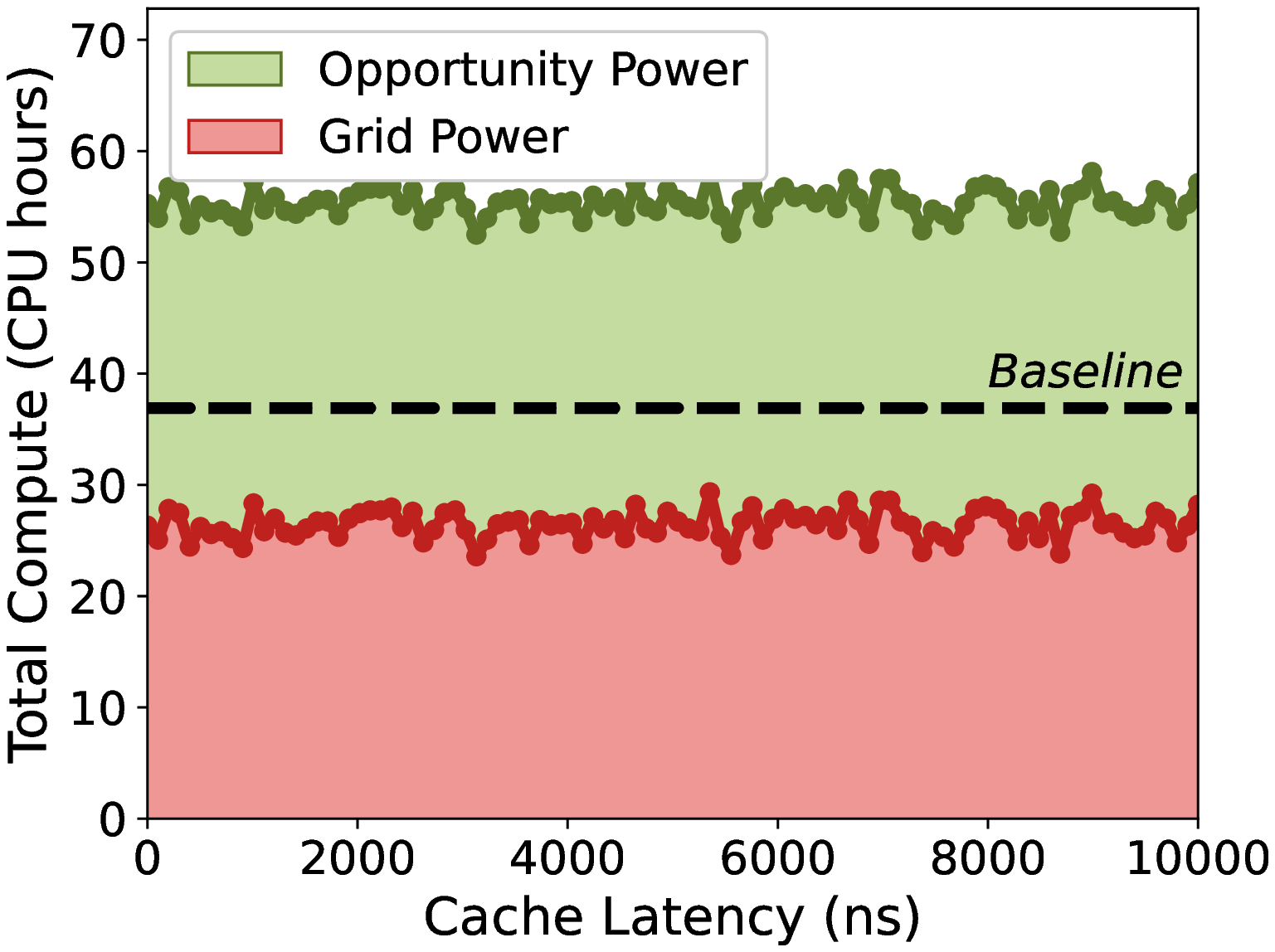}} 
\vskip 1em
\subfloat[Small (100ms) Job]{\includegraphics[width=5cm]{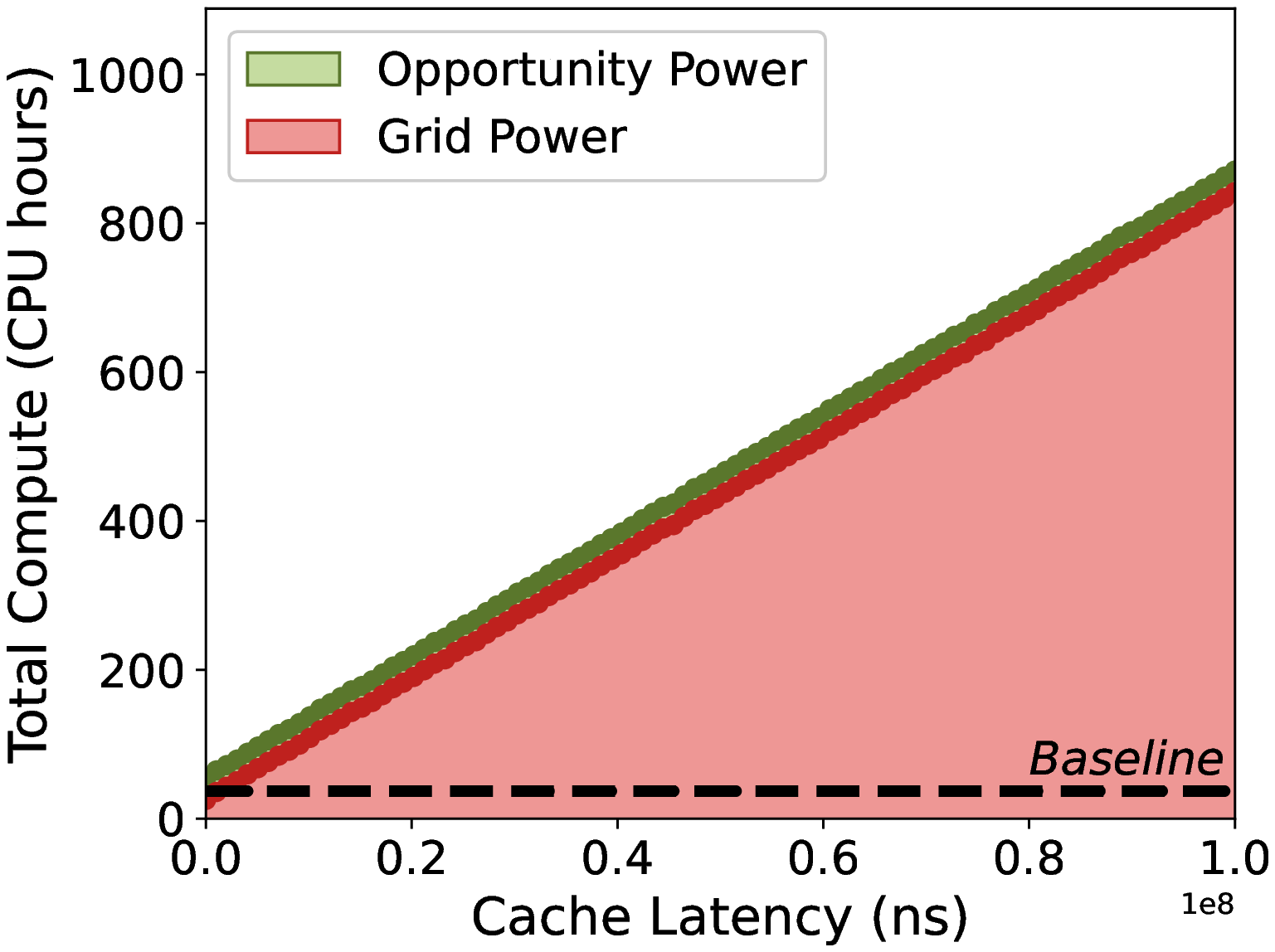}}\hfil   
\subfloat[Medium (10s) Job]{\includegraphics[width=5cm]{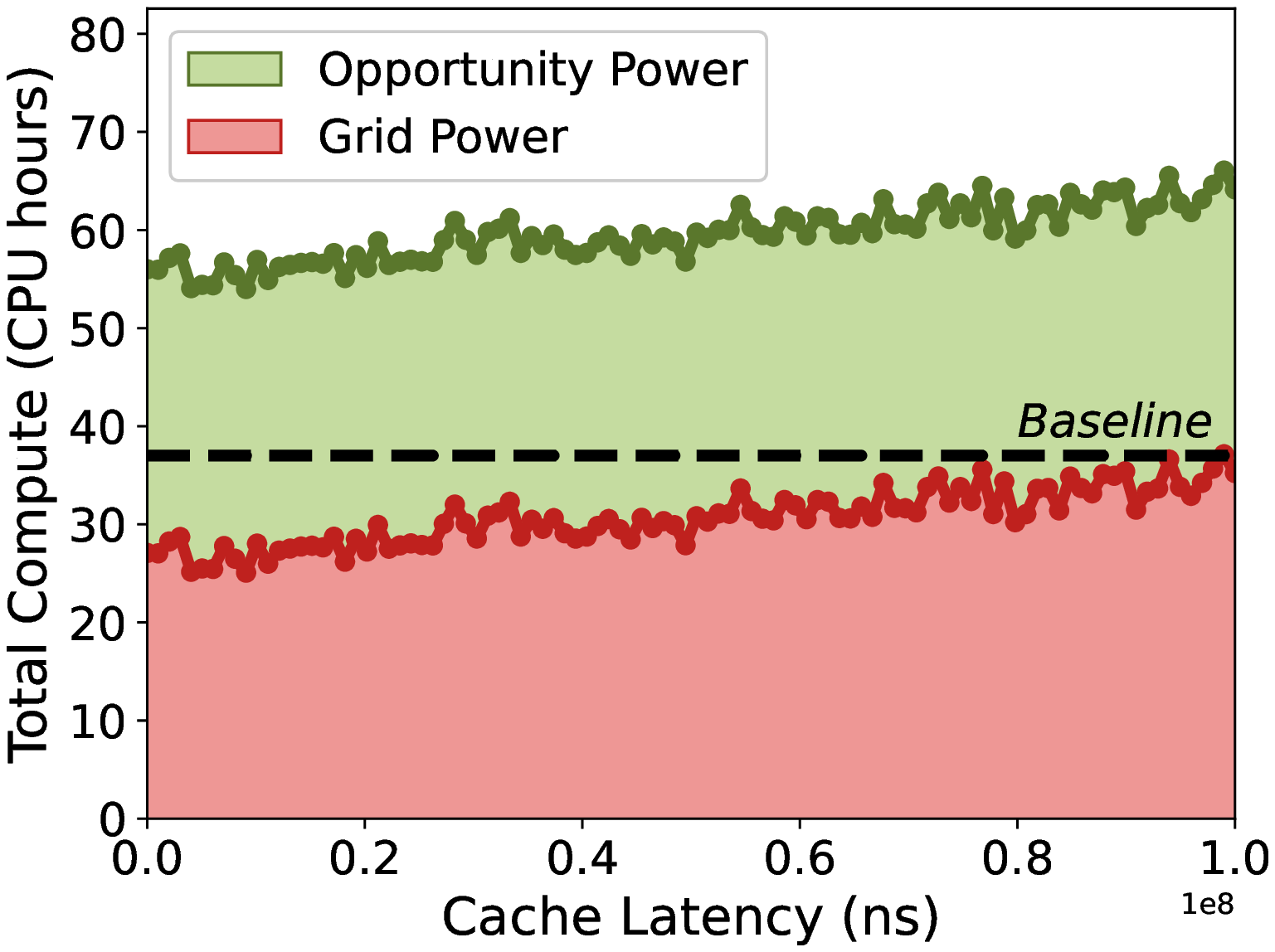}}\hfil
\subfloat[Large (60s) Job]{\includegraphics[width=5cm]{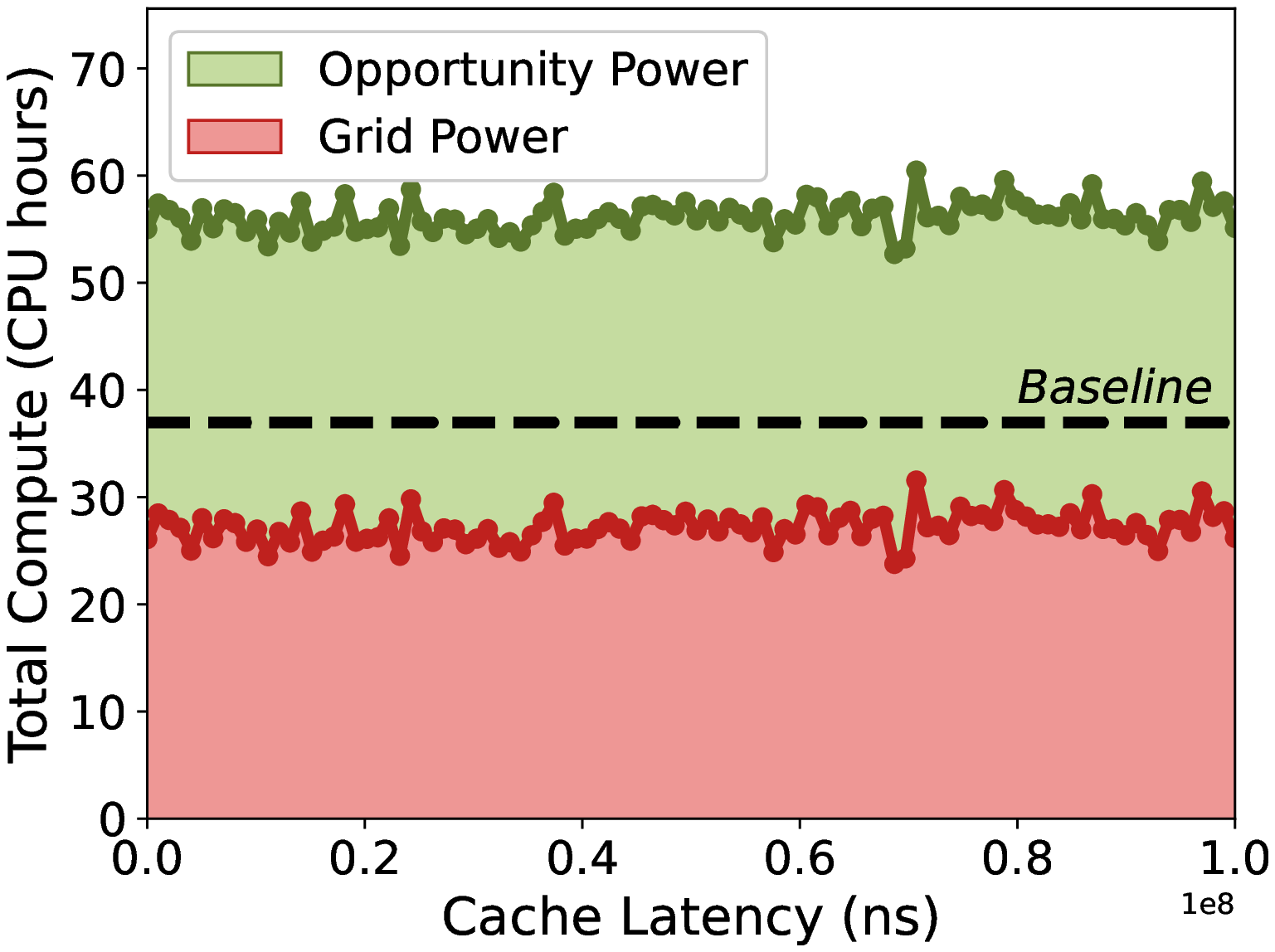}}
\caption{\textbf{Evaluating the effect of cache latency on energy savings.} Results presented for small (100ms), medium (10s), and large (60s) jobs. Cache latency has a huge effect on performance for short jobs.} \label{figure}
\end{figure*}

Our initial simulation results highlight the importance of accurate price prediction (Figure~\ref{fig:fp}), and low cache latency (Figure~\ref{figure}), and relatively accurate task prediction (Figure~\ref{fig:caching}). We use the results from microbenchmarking to realistically parameterize our system-level simulator according to the metrics described in Table~\ref{tab:sim}.

We then evaluate the overall effectiveness of Information Batteries according to: 1) the efficiency of the system, in terms of the amount of processing offloaded from grid power to opportunity power, and 2) dollar cost relative to traditional batteries.

\subsubsection{Energy savings.}
A key goal of the IB system is to make use of opportunity power that might otherwise be wasted. Thus, a central metric for any IB system is the amount of processing power offloaded from non-opportunity power to opportunity power.

We measure these savings in terms of the number of machine cycles offloaded. Figure~\ref{fig:caching} shows these results for a wind-dominant power profile (MISO). We parameterize the simulator as follows: (1) Memoization overhead of 1 ms (2) Task prediction accuracy of 50\%; (3) For the price predictor, false positive and false negative rates of 0.1\%. The length of tasks is varied.

\begin{table}
\begin{tabular}{ll}
\hline
    \textbf{Cache Miss} &  1348ns\\
    \textbf{Cache Hit (Fetch)} & 1455ns\\
    \textbf{Store} & 2262273ns\\
\hline
\end{tabular}
\vskip 0.5em
\caption{Average latency of cache hit, cache miss, and store.}
\label{tab:mem}
\vskip -1.5em
\end{table}

\begin{table}
\begin{tabular}{lll}
    & Mean-Absolute & Mean-Absolute \\
    & Error (Val) & Error (Train) \\ \hline
CAISO & 0.1181 & 0.1402 \\
MISO & 0.0745 & 0.0614 \\
\hline
\end{tabular}
\vskip 0.5em
\caption{Performance of the price predictor model on validation and training data. The model performed best on MISO data, but the mean-absolute error was low for both datasets.}
\label{tab:model_results}
\vskip -1.5em
\end{table}

Note that Information Batteries always incur more machine cycles than traditional compute. This is to be expected, since the IB system must perform the same computational task with additional pre- and post-processing. However, in these scenarios, the grid power consumed by the IB system is less than that for traditional compute, since a large fraction of the compute was offloaded to opportunity power.

The overall energy savings of the IB system is the difference between the grid power consumed in the traditional compute scenario and the grid power consumed by the information batteries.

\subsection{Cost of storage}
In some scenarios, Information Batteries are more cost-effective than traditional energy storage systems.
First we explain how to think about the storage capacity of an IB system. Then we compare potential IB systems with conventional grid-scale battery storage.

Unlike conventional batteries, Information Batteries have both a ``charge rate`` measured in Watts, as in the power consumption of the data center---corresponding to the power draw rate at which precomputation can be done by a given IB system---and a second dimension not present in ordinary batteries corresponding to a prediction time horizon.  In some ways this makes an IB system incommensurable with conventional batteries.

\subsubsection{IB memoization overhead.}
Our function-level precomputation has an overhead of 368$\mu$s for each put, 10$\mu$s for each cache miss and 7$\mu$s for each cache hit and retrieval on a 2.6 GHz Intel Core i7 CPU. This is extraordinarily efficient by virtue of its simplicity. For example, any job with just one second or longer run time would experience less than 1\% overhead due to memoization.

\subsubsection{Battery comparison.}
Consider for a moment a hypothetical IB system that has a one day time horizon, and can predict with perfect accuracy.  The IB system can thus precompute a day's worth of tasks. Consider a hyper-scale data center that is 100 MW with this one-day prediction horizon; an IB system in this data center could store a monumental 8.64 TJ via precomputation alone.

Of course, it is rare for a data center to have full-day lookahead.  Instead, we might more realistically have, on average, 90 minutes prediction ability with 90\% accuracy.\footnote{There is little public data on workloads and scheduling; our estimate here is based upon our experience working in such hyper-scale compute environments.}  With one-hour lookahead, such a data center could store 540 GJ, significantly more than most grid-scale battery-based storage projects.  Given the negligible overhead of memoization, the key efficiency parameter is job prediction.

Using 90\% as a canonical target efficiency, as it closely matches lithium-ion battery efficiencies, such an IB-enabled data center would match the storage capacity of a 150,000 kWh lithium-ion storage array, which, at current lithium-ion prices of \$356 / kWh~\cite{lion-cost} would cost \$53.4 million.

%% file: discussion.tex
\section{Conclusion}
In this work we have shown that Information Batteries have the potential to provide a cost-effective means to cope with growing renewable intermittency using large-scale computing infrastructure.  Key to the IB approach is that it is not a general-purpose solution, but is likely to be effective for many common workloads. 

\subsection{Future directions}
This paper merely introduces and explores one avenue of implementation and evaluation of Information Batteries.  Much remains to be studied.  In particular, we believe that there are three lines of worthwhile future research on this topic: improved prediction, improved integration into large systems, and support for the precomputation and recombination of fragments of computational tasks.

\subsubsection{Prediction.} The smart grid research community has done extensive work to improve prediction of price and power availability.  In addition, the distributed systems community has done extensive work to characterize workloads in a wide range of settings. As we showed, as prediction accuracy improves, the efficiency of an IB system will improve, so there is substantial low-hanging fruit in incorporating state-of-the-art predictors.

\subsubsection{Integration.} While we frame our IB system prototype as an end-to-end system, any real-world deployment of this approach would necessarily omit the toy compute controllers we built for testing and instead integrate IB decision-making into an existing compute controller (e.g., a Kubernetes controller managing a whole data center).  The criteria used by such data center operators to use the IB approach would necessarily be dependent upon their costs, business models, and the types of workloads they typically serve. 

\subsubsection{Computation.} The efficiency of an IB system depends in large part on how well jobs can be accurately predicted and precomputed.  But this precomputation need not be merely binary in nature---indicating whether a whole task should be precomputed or not---but instead can reflect a complex planning strategy that decomposes compute workloads into precomputable sub-units.  There remains substantial work to be done on integrating ideas from related analyses performed in dramatically-different contexts, such as return-oriented programming, to identify which tasks can be meaningfully fragmented and then reassembled. In addition, efficient caching and retrieval of fragmented, precomputed resulsts is challenging, as the greater complexity of fragmented precomputation, the more expensive retrieval is likely to be; this may require storage of program control-flow graphs along with compute fragments, so as to easily identify cached results that will meet the needs of new tasks.  Finally, our exploration in this paper leveraged compiler support for program instrumentation, but in a real deployment it would be ideal to support unmodified program binaries.